\documentclass[10pt,a4paper,twocolumn]{article}
\usepackage[utf8]{inputenc}
\usepackage{amsmath}
\usepackage{amsfonts}
\usepackage{amssymb}
\usepackage{times}
\usepackage[small]{titlesec}
\usepackage{graphicx}
\usepackage{microtype}
\usepackage[top=0.7in, bottom=1in, left=0.8in, right=0.8in]{geometry}
\usepackage{titling}

\usepackage{placeins} %Pone la tabla donde uno quiere
\usepackage{float}
\restylefloat{table}

\title{Uncertainties in permittivities computed from molecular dynamics simulations and temperature correction of dielectric properties of condensed polar systems}

\date{\small
	$^1$Instituto de F\'{\i}sica de L\'{\i}quidos y Sistemas Biol\'{o}gicos. CONICET  La Plata - UNLP, Argentina.\\%
	$^2$Instituto de Ingeniería y Agronomía, UNAJ. Florencio Varela, Buenos Aires, Argentina.\\
    $^3$Universidad Tecnol\'{o}gica Nacional - FRBA, UDB F\'{\i}sica, Buenos Aires, Argentina.\\[2ex]%
	\today
}

\begin{document}

\author{Hernán R. Sánchez$^1$, Ramiro M. Irastorza$^{1,2}$ and C. Manuel Carlevaro $^{1,3}$ }

%\affil[1]{Instituto de F\'{\i}sica de L\'{\i}quidos y Sistemas Biol\'{o}gicos
%(IFLYSIB)-CONICET- CCT La Plata, Argentina.}
%\affil[2] {Instituto de Ingeniería y Agronomía, UNAJ. Florencio Varela, Buenos Aires, Argentina.}
%\affil[3] {Universidad Tecnol\'{o}gica Nacional - FRBA, UDB F\'{\i}sica, Mozart No 2300, C1407IVT Buenos Aires, Argentina.}

\twocolumn[
	\begin{@twocolumnfalse}
		\maketitle		
		\begin{abstract}
        A robust, simple and fast procedure for the calculation of uncertainties in relative static dielectric permittivity ($\varepsilon_s$) computed via molecular dynamics (MD) is proposed.  It arises as a direct application of well founded statistical methods for auto-correlated variables. Also, in order to deal with the lack of experimental data about $\varepsilon_s$ and relaxation times ($\tau$) at different temperatures, a method for their prediction is suggested. It requires one experimental value and at least two MD simulations. In the case of relaxation times, a theoretical justification is provided.
		\end{abstract}
	\vspace{1cm}
	\end{@twocolumnfalse}
]

\section{Introduction}	

The dielectric relaxation processes of liquids have historically measured and well documented in literature \cite{shinomiya1989dielectric,gregory2001tables}. One reason for this is that liquids are frequently used as calibration media of experimental setups. More interestingly, the behavior of hydrogen bonding of macromolecules (e.g.: biopolymers) can be infered by dielectric properties measurements \cite{kaiser2016hydrogen}. These properties are determined by the interaction of external electromagnetic fields with matter and 
are extensively used in a variety of applications. However, at molecular level these interactions are poorly understood and still constitutes a very active research area \cite{english2015perspectives}. Molecular dynamics (MD) can be used for the computation of dielectric properties, but they provide some of the greatest challenges  among of the magnitudes commonly obtained by this method \cite{caleman2011force}. Computed values are not accurate enough to be considered quantitative predictions, and do not improve upon experimental values measured at relatively close temperatures. 

Among many empirical parameterizations of the dielectric properties of liquids \cite{hilfer2002h}, this paper will study the simple Debye relaxation model, as far as we understand, without loss of generalization in the analysis. Such a exponential model needs three parameters, we will focus on relative static dielectric permittivity ($\varepsilon_s$) and relaxation time ($\tau$). These parameters have not been measured at different temperatures for most common chemical species, notwithstanding the above, usually their temperature dependence is not negligible. In particular, relaxation times can easily vary in factor 5 with a few tens of kelvins \cite{gregory2001tables}. It is therefore evident the importance of having a method for predicting those magnitudes.

In this work we propose a simple MD based method to predict those magnitudes at some temperature, which requires knowing the corresponding experimental value at some reference temperature.

Previously we address the following issue. If several MD simulations are started with different initial conditions, different values of permittivity are obtained. Then, permittivity can be treated as a random variable. As such, some measure of its dispersion is of enormous value. Classical equations for the variance of the mean cannot be directly employed. In this work, we describe and advocate the use of a robust methodology for this purpose.

\section{Methods}

\subsection{Calculation of dielectric properties}
If the hypothetical fields involved are small enough, the microscopic fluctuations of a system at thermodynamic equilibrium in absence of external fields determine the properties under study. % Sino hay problemas por no ser lineal

% ALTERNATIVA Despite dielectric properties are closely connected with responses induced by fields, they can be obtained from equilibrium properties in absence of external fields.

The relative static dielectric permittivity ($\varepsilon_s$) of polar systems can be obtained through the dipole moment ($\mathbf{M}$) of the system in absence of external fields. In IS units, they are related by\cite{neumann1983dipole}

\begin{equation}\label{eq: permit}
\varepsilon_s = \frac{E[\mathbf M^2] - E[\mathbf M ]^2 }{3\varepsilon_0 VT} + 1
\end{equation}

\noindent where $E[\cdot]$ stands for expected value,  $\varepsilon_0$ represents the permittivity of free space and, $V$ and $T$ the expected value of the system volume and temperature, respectively. We omitted the expected values for simplicity.

By definition, numerator in Eq. \ref{eq: permit} is the variance of $\mathbf{M}$. Due to the finite nature of practical MD simulations, its value cannot be obtained but estimated. However, it is known that $E[\mathbf M ] = 0$ by symmetry. This knowledge let us to use a better estimate of the population variance\cite{zhang1996estimating} which turns to be an estimator of $E[\mathbf M^2]$. Further details can be found in Section \ref{section: uncertainties}.

Relaxation times can be obtained from the time auto-correlation function ($\phi(t)$) of the total dipole moment ($\mathbf{M}(t)$)\cite{gordon1968correlation}. It is defined by

\begin{equation*}
\phi(t) = \frac{\langle \mathbf M(t^\prime) \, \mathbf M(t-t^\prime) \rangle }{\langle \mathbf M^2(t^\prime) \rangle }
\end{equation*}

This function allows modeling the relaxation process in time domain. The parameters corresponding to well known relaxation models can be estimated by fitting the autocorrelation function to the corresponding expression. For present purposes the Debye model\cite{debye1912einige} was chosen, which in time domain can be represented by

\begin{equation}\label{ec: fx debye}
f(t) = e^{-t/\tau_D}
\end{equation}

It has the benefit of simplifying the comparison processes due to it has only one parameter, i.e. the relaxation time $\tau_D$. Despite models that generalize beyond the previous one\cite{hilfer2002h} fit at least equally good to experimental data, errors due to simulation inaccuracies are orders of magnitude larger.

\subsection{Computation of uncertainties\label{section: uncertainties}}
\subsubsection{Uncertainties of $\boldsymbol{\varepsilon_s}$}

Relative static dielectric permittivity can be computed using Eq. \ref{eq: permit}. The uncertainty in $\varepsilon_s$ computed through Eq. \ref{eq: permit} comes almost exclusively from its numerator, which correspond to the variance of the total dipole moment ($\sigma^2$). Then $T$ and $V$, the estimates of the expected value of temperature and volume, can be considered constant values.

The key point of our contribution is to recognize that the dispersion in the relative static dielectric permittivity can be estimated computing the variance of an estimator of the variance of the total dipole moment of the system, being the latter  an auto-correlated variable.

The dispersion of $\varepsilon_s$ can be quantified with its variance ($\text{Var}(\varepsilon_s)$). If  $\hat{\sigma}^2$ is an estimator of $\sigma^2$,

\begin{equation*}
\text{Var}(\varepsilon_s)\approx  \text{Var}(\hat{\sigma} ^2)/(3\varepsilon_0 VT)^2
\end{equation*}

The values of $\mathbf M(t)$  are auto-correlated which implies that classical variance estimators are not unbiased but only asymptotically unbiased. Although unbiased estimators can be used, the bias of well known estimators can be neglected if the MD simulation has a typical length. Nevertheless, we need to use proper expressions for the variance of the variance estimators.

Above mentioned expressions depend upon auto-correlation function. The true value remains unknown but many estimators can be used. This, along with a series of approximations give raise to a series of estimators. In our experience, they are similar enough to make their individual analysis outside the scope of this work. Among many alternatives, here we only describe the used for the results presented below.

We employed the most common discrete estimator of the auto-correlation function. For $n$ registered steps, evenly-spaced at time intervals of duration $\xi$, it is defined by

\begin{equation}
r(k\xi) = \frac{\sum_{i=1}^{n-k} (\mathbf M(i\xi)-\bar{\mathbf M})(\mathbf M((i+k)\xi)-\bar{ \mathbf M}) }{\sum_{i=1}^{n} (\mathbf M(i\xi)-\bar{ \mathbf M})^2}
\end{equation}

\noindent where $\bar{\mathbf M}$ is the sample mean of the total dipole moment. It is used to construct an estimator of the effective number of observations, $n_{\text{eff}}$, which is defined by\cite{bayley1946effective}

\begin{equation*}
n_{\text{\text{eff}}}= \frac{\text{Var}(x)}{\text{Var}(\bar x)}
\end{equation*}

\noindent for some auto-correlated random variable $x$.
%\begin{equation*}
%n_{\text{eff}}= \frac{n}{1+ 2 \sum_{k=1}^{n-1}\left( 1-k/n\right)\phi(k\,\xi)}
%\end{equation*}

The estimator that we chose is\cite{zikeba2011standard}

\begin{equation}
\hat n_{\text{eff}} = \frac{n}{1+ 2 \sum_{k=1}^{n_c}\left( 1-k/n\right)r(k\,\xi)} 
\end{equation}

\noindent where for the limiting lag, $n_c$, we used the one corresponding to the first transit of the auto-correlation function
estimate through zero method, that is $n_c=\min \{ k|r_k>0\wedge r_{k+1}<0 \}$. It can be shown that $n_c$ exist even if $\phi(k\xi)>0\forall k$\cite{percival1993three} as in our case.

The sample variance of an autocorrelated variable ($s_{\text{AC}}^2)$ satisfy $s_{\text{AC}}^2 = s^2 n_{\text{eff}}/(n_{\text{eff}}-1)$, where $s^2$ is the commonly used estimator of the variance of uncorrelated variables. For the computation, $n_{\text{eff}}$ can be replaced by $\hat n_{\text{eff}}$. When simulation is many times longer than relaxation time, $n_{\text{eff}} $ is large and $s_{\text{AC}} \approx s$.

Finally, the variance of $s_{\text{AC}}^2$ satisfy

\begin{equation}\label{ec: var s2}
\text{Var}(s_{\text{AC}}^2) = \frac{2\sigma^2}{\nu_{\text{eff}}}
\end{equation}

\noindent where the effective degrees of freedom ($\nu_{\text{eff}}$) can be asymptotically approximated according to

\begin{equation}
\nu_{\text{eff}} \cong \frac{n}{1 + 2\sum_{k=1}^{n-1}\phi(k\, \xi)^2}-1
\end{equation}

For the computation, we used the following estimator

\begin{equation}\label{eq: nu eff estimation}
\hat \nu_{\text{eff}} = \frac{n}{1 + 2\sum_{k=1}^{n_c}r(k\, \xi)^2}-1
\end{equation}

If the simulation is large enough,  $\sigma$ can be replaced by $s_{\text{AC}}$ in Eq. \ref{ec: var s2}, and we get

\begin{equation}\label{eq: proposed var e_s calculation}
	\text{Var}(\varepsilon_s) =  \frac{2 s_{\text{AC}}^2}{\hat \nu_{\text{eff}} \, 9\varepsilon_0^2 V^2 T^2}
\end{equation}

\subsubsection{Uncertainties of relaxation times}
Computed relaxation times are random variables because of the nature of the process used to obtain them. It is desirable to provide some estimation of their uncertainty. They are closely related to the values of relaxations times as showed bellow.

At first, it is worth observing that not all lags are used for fitting the auto-correlation function. Uncertainty in the auto-correlation function estimates increase with lags, and this may creates oscillations large enough to be considered as noise. Because of this, in practice we only considered those lags representing times shorter than few $\tau_D$. This strategy is similar to the employed in the first transit through zero method. The variance of the $k-$th lag is given by\cite{andersonbook}

\begin{equation}
\text{Var}(r_k) \simeq \frac{1}{N} (1+2\sum_{i=1}^{K}r_i ^2)
\end{equation}

\noindent where $K<k$, and $N$ represent the number of samples. For the present cases, the auto-correlation function and its estimates can be approximated by Eq. \ref{ec: fx debye}. This imply that the summation above, and consequently $\text{Var}(r_k)$ is approximately proportional to $\tau_D$ for common values of $N$. It can be seen by integrating the square of the auto-correlation function, estimated samples of which are included in the summation above, between zero and a fixed constant multiplied by $\tau_D$. By these means we find that the uncertainty in $\tau_D$ is proportional to $\tau_D$ up to first order. Nevertheless, as $\tau_D$ belong to an exponential function, we can only claim that the uncertainty in $\tau_D$ tends to increase with $\tau_D$.

%We have shown that the relaxation times predicted only from simulation results are appropriate to be used together with the proposed method. However, they are random variables because of the nature of the process used to obtain them, and we would like to provide some estimation of their uncertainty.  

%To this end, we performed three longer simulations for TIP4P at 50$^\circ$ C under the same remaining conditions, resulting in three additional samples of 400 ns each one. The selection criteria was to get a simulation that minimize the product between relaxation time and computational cost. We split the total run in 30 parts of the same length that our previous simulations, and computed the corresponding relaxation times. Their standard deviation was 0.061 ps. This is an estimate of the uncertainty for the case in question.

% We can provide a rough estimation of the uncertainties of the remaining cases. 

%So, up to first order, the uncertainty in $\tau_D$ is proportional to $\tau_D$, and we can use the value for TIP4P at 50$^\circ$ C for making prediction of the uncertainties of the remaining cases. They can be found in table \ref{tablatau} under columns labeled as SD.

% https://ned.ipac.caltech.edu/level5/Stetson/Stetson1_3_3.html
% https://stats.stackexchange.com/questions/59555/standard-errors-of-regression-coefficients-based-on-sample-size

\subsection{Temperature dependence}
In order to predict the temperature variation of relaxation time we need to consider some model. Following Eyring ideas\cite{eyring1936viscosity}, the relaxation process taking place in a system of dipoles can be understood as an activated process, and the corresponding relaxation time approximately obeys\cite{kauzmann1942dielectric}

\begin{equation}\label{Modelo activado tau}
\log \tau(T) = \frac{\Delta H}{RT} - \log \frac{h}{kT} - \frac{\Delta S}{R}
\end{equation}

\noindent where $\Delta H$ and $\Delta S$ represent the barrier enthalpy and entropy, respectively. $T$ represents the equilibrium system temperature,  and $R$, $h$ and $k$ the gas, Plack and Boltzmann constants, respectively.

If experimental ($\mathcal{E}$) and simulated ($\mathcal{S}$) data follow equation (\ref{Modelo activado tau})  the relaxation times approximately satisfy
%We neglect the temperature dependence of the thermodynamics functions, as discussed bellow. By doing so and
\begin{equation*}
\tau_{\mathcal{E}}(T_2) = \tau_{\mathcal{E}}(T_1) \frac{\tau_{\mathcal{S}}(T_2)}{\tau_{\mathcal{S}}(T_1)} e^{\frac{[\Delta H_\mathcal{E}-\Delta H_\mathcal{S}](T_1-T_2)}{RT_1T_2}}
\end{equation*}

We made some assumptions while deriving the equation above. The entropy variation with temperature of condensed matter systems can be approximated by the well known expression $C_p \ln (T_2/T_1)$, where $C_p$ is the heat capacity at constant pressure, and it was considered independent of the temperature in the interest range. If we take the difference  $\tau_{\mathcal{E}}(T_2)-\tau_{\mathcal{E}}(T_1)$, the terms involving entropy can be expressed as the difference between the increase in entropy with temperature for the activated state and the increase in entropy with temperature for the relaxed state, both divided by $R$. This can be rewritten as $(C_{p,\text{activated}}-C_{p,\text{relaxed}} )R^{-1}\ln (T_2/T_1)$. The difference in parenthesis is very small, as it can be realized as the variation in heat capacity due to the influence of a small electric field. The remaining multiplicative factor is also small for reasonable temperature ranges. The same can be done for $\mathcal{S}$, therefore we can avoid the terms containing the entropy variable.

$\Delta H_\mathcal{E}$ cannot be computed from only one experimental value. Being that the process consists in reorient dipoles, the absolute value of both $\Delta H$ is expected to be small, therefore we further approximate that the argument of the exponential goes to zero.

Despite some systems cannot be accurately characterized by only one relaxation time, there is normally one main relaxation time, in the sense that a pronounced relaxation takes place around that time.

% EVENTUALMENTE COMENTAR LO DE CALCULAR PRIMERO LA ENTROPÍA Y...

\subsubsection{Systems under study}
The selection of systems for simulation was restricted to those ones for which experimental data were available for comparison.
%The choice of systems for simulation is restricted to those ones for which there are experimental data available to compare with.

We chose six substances in liquid state: dimethyl sulfoxide (DMSO), water, methanol, ethanol, propanol and buthanol. In the case of water, the experimental values were taken from references \cite{ellison2007permittivity,kaatze1989complex}. The remainder values come from reference \cite{gregory2001tables}. We fitted the Debye parameters when not reported. 

According to availability, we consider data at 10, 20, 30, 40 and 50 $^\circ$C, except for DMSO for which the value at 10$^\circ$C is not available. Final values can be found in the tables \ref{tablaerroreps} and  \ref{tablatau}.

\subsubsection{Simulation details and analysis}
The software Gromacs 2016.4\cite{abraham2015gromacs} was employed for the simulations. They were performed in cubic boxes with no external fields applied. We used all-atoms molecular models with no additional geometrical  constraints. The initial geometries were generated with Packmol\cite{martinez2009packmol}. Every system contains in 1000 molecules, this number is considered reasonable for this kind of calculations\cite{olmi2016can}. Long range coulombic interactions were modeled  with reaction field[CITA]. The elected force field was OPLS\cite{jorgensen1988opls,jorgensen1996development} and the TIP4p model\cite{jorgensen1983comparison} was used for water.

The following procedure was performed for each compound: Firstly, the system energy was minimized. Then they were simulated for 42 ns in NpT conditions at the lowest considered temperature for the compound. The first 2 ns were exclusively used for equilibration. We employed the Berendsen's barostat  which does not adversely affect the predicted values for the interest properties despite it does not provide truly canonical trajectories\cite{kaiser2016hydrogen}. In fact, they seem to be rather insensitive to pressure as previously NVT values were  indistinguishable from those from NpT simulation\cite{kaiser2016hydrogen}. The simulation equilibrium pressure was set to 1 bar. The temperature was subjected to the Bussi-Donadio-Parrinello velocity rescaling algorithm\cite{bussi2007canonical}. Initial volumes were chosen according to experimental values. Time step were set up at 2 fs for water and DMSO, and 1 fs in other cases.

The final geometries were the starting point for subsequent runs performed at the remaining temperatures and the same simulation conditions. In all cases, the initial velocities of the molecules were randomly generated.

We develop our own Python3\cite{van2014python} routines based on the MDTraj library\cite{mcgibbon2015mdtraj}, and used them for analyzing the simulation trajectories.

\section{Results and discussion}

\subsection{Estimation of uncertainties of $\boldsymbol{\varepsilon_s}$}
Given a simulation, the suggested method allows to directly compute the uncertainty in $\varepsilon_s$. %Alternatively, it can be done using the following procedure which is more intuitive:  %
Alternatively, the following more intuitive procedure can be used:
at first the simulation is evenly split in $m$ contiguous parts, for each part $\varepsilon_s$ is computed and the standard deviation of those values ($s_{\varepsilon_s}(m)$) is calculated. From equations \ref{eq: nu eff estimation} and \ref{eq: proposed var e_s calculation} follows that if sufficient number of samples are included in the analysis, the standard deviation of $\varepsilon_s$, for a given substance at some temperature, is almost proportional to the inverse of the square root of the number of samples ($n^{-\frac{1}{2}}$). In brief,  it is expected for $s_{\varepsilon_s}(m)$  to be almost proportional to $\sqrt{m}$.

%As standard deviation of the mean is proportional to the the square root of the number of partitions used ($m$), it is expected for $s_{\varepsilon_s}(m)$  to be proportional to $\sqrt{m}$, as the variance of $\varepsilon_s$ is approximately inversely proportional to the number of samples REDUNDANTE/CONFUSO. 

In order to obtain a numerical approximation of $s_{\varepsilon_s}$ for the entire simulation, we performed the corresponding computations for many values of $m$, fit the $s_{\varepsilon_s}(m)$ to a linear function, and extrapolated to $m=1$. 

The fitting procedure was carried out using ordinary least squares. We used a linear model without intercept. The addition of this parameter to the model leads to unstable results, and in some cases might even leads to negative estimates. For the numerical experiments, we used  $\{m\in\mathbb{N}| 1 < m \leq 100\}$. 

\begin{figure}[tbph!]
	\centering
	\includegraphics[width=1.0\linewidth]{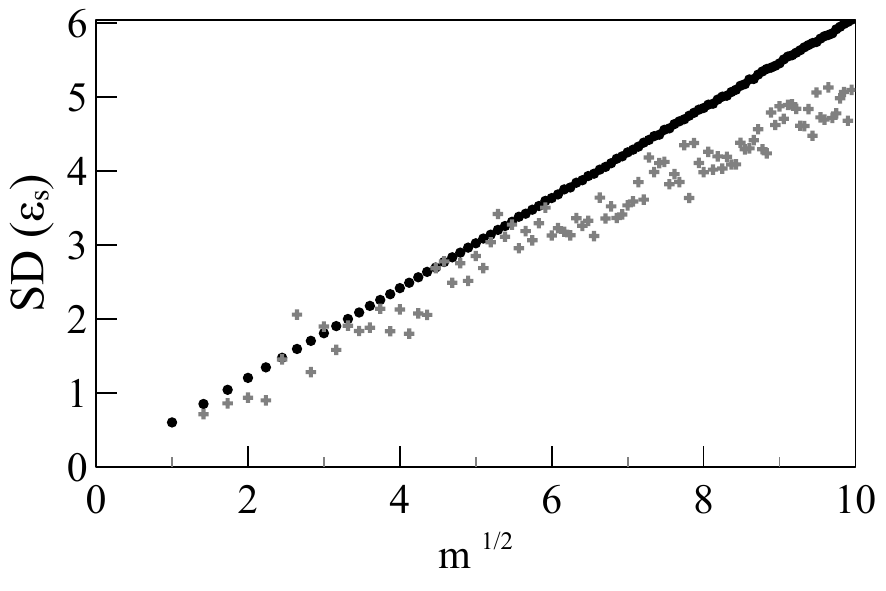}
	\caption[Leyenda corta]{Predicted standard deviation of $\varepsilon_s$ vs. square root of the number of parts in which the simulation (TIP4P at 30$^\circ$C) was divided. The proposed and numerical method are represented by black dots and gray crosses, respectively. }
	\label{fig:erroresepsilones}
\end{figure}

For comparative purposes, the same partition scheme and procedure was used in conjunction with the suggested method, except that mean was used instead of standard deviation. This is due to an estimate of the standard deviation was computed for each part. Also, we included the case $m=1$. For illustrative purposes, we show the values for TIP4P at 30$^\circ$C  in Figure \ref{fig:erroresepsilones}

The robustness of the proposed procedure is seen by simple inspection. Our hypothesis of linearity was corroborated for TIP4P at 30$^\circ$C for which we obtained a determination coefficient of 0.999994. There is a lost in linearity as $m$ and $\tau$ increase. However, the linearity is seen even for simulations that are not long enough to be used for predicting this property with acceptable accuracy. This suggest that the proposed method works even for simulation times many times shorter than those required for quantitative purposes.

In table \ref{tablaerroreps} we included the standard deviations predicted through the procedures above. The standard deviation of the coefficient was omitted because the employed partition scheme imply mutual dependency of the errors.

\begin{figure*}[tbph!]
	\centering
	\includegraphics[width=0.9\linewidth]{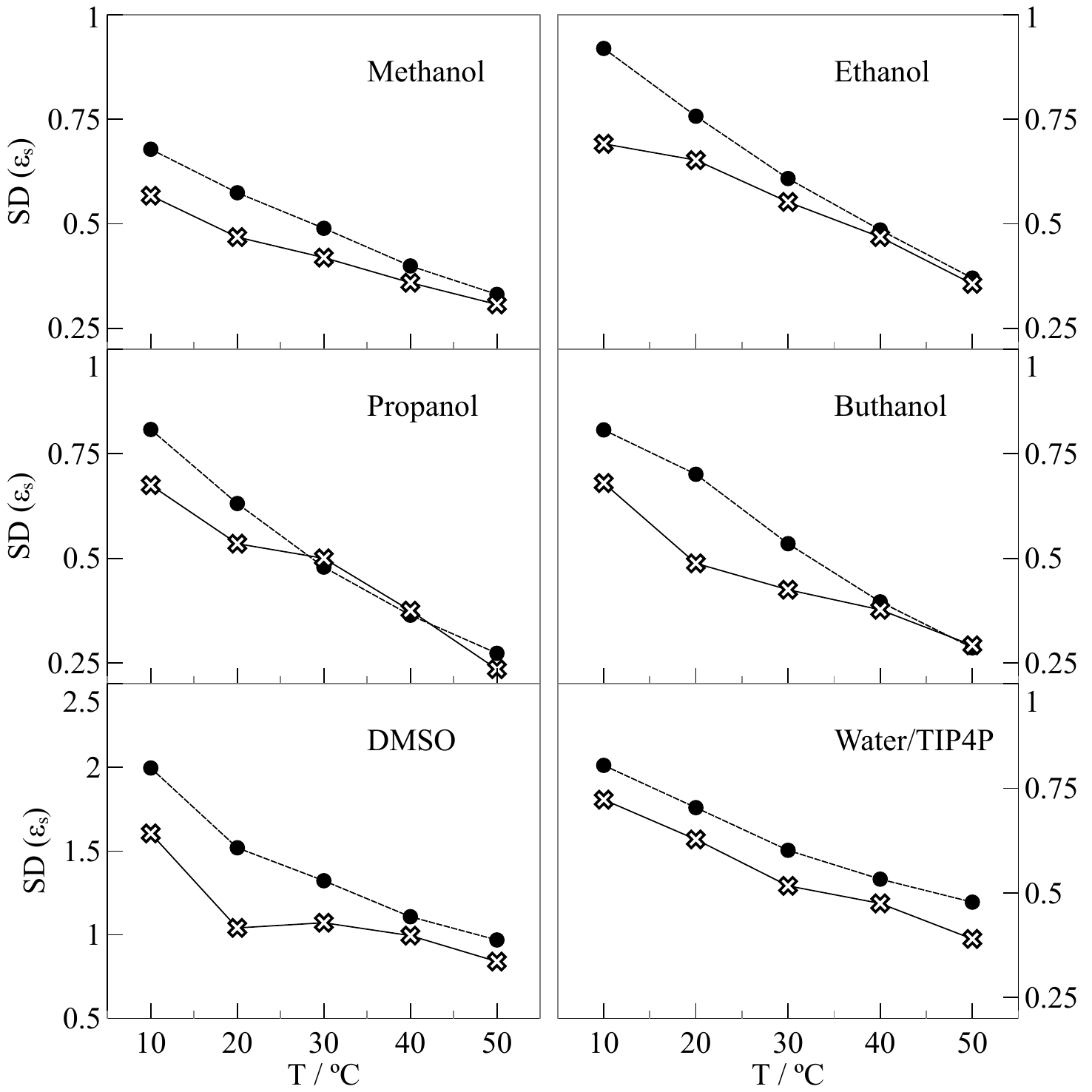}
	\caption[Leyenda corta]{Predicted standard deviation of $\varepsilon_s$ vs. temperature (in K) for the six substances. The proposed and numerical method are represented by dots and crosses, respectively. }
	\label{fig:erroresepsilonestemperatura}
\end{figure*}

Although numerical estimations are pretty acceptable, they do not present a smooth temperature dependence as the proposed approach does, which is clearly evident in the DMSO case, please see Figure \ref{fig:erroresepsilonestemperatura}. This suggest that proposed method provides very consistent results, and that it is more robust than the more intuitive approach, which can be understood as an application of slightly modified block averaging method\cite{newman1999monte}. Also, notice that the computational cost of our approach is several orders of magnitude less that the required for alternative methods like bootstrap or jackknife\cite{newman1999monte}.

\begin{table*}[tbph!]%[htb]
	\centering
	\caption{Magnitudes related to $\varepsilon_s$. $^a$ Raw MD results. $^b$ Values predicted with the proposed method. $^c$ Experimental values. $^d$ Standard deviation computed through the numerical approach. $^d$ Standard deviation computed through the proposed method. \label{tablaerroreps}}
	\begin{tabular}{ccccccccccc}
		\hline
		T($^\circ$ C) &Sim.$^a$& Pred.$^b$& Exp.$^c$& SD num.$^d$& SD prop$^e$.  &Sim.$^a$& Pred.$^b$& Exp.$^c$& SD num.$^d$& SD prop$^e$.   \\
		\rule[-1ex]{0pt}{3.8ex}&& &{\small\textbf{Methanol}}& & & &&{\small\textbf{Ethanol}}&& \\
		10 & 27.55  & 35.74 & 35.74  & 0.567 & 0.678  & 20.66 & 26.79  & 26.79 & 0.691  & 0.919 \\  
		20 & 26.40  & 33.79 & 33.64  & 0.468 & 0.574  & 19.42 & 25.22  & 25.16 & 0.652  & 0.757 \\  
		30 & 25.02  & 31.84 & 31.69  & 0.419 & 0.489  & 18.37 & 23.66  & 23.65 & 0.552  & 0.608 \\  
		40 & 23.20  & 29.88 & 29.85  & 0.359 & 0.399  & 17.24 & 22.09  & 22.16 & 0.468  & 0.485 \\  
		50 & 21.72  & 27.92 & 28.19  & 0.307 & 0.331  & 15.82 & 20.52  & 20.78 & 0.356  & 0.370 \\
		\rule[-1ex]{0pt}{3.8ex}&& &{\small\textbf{Propanol}}& & & &&{\small\textbf{Buthanol}}&& \\
		10 & 14.92  & 22.61 & 22.61  & 0.675 & 0.808  & 11.83 & 19.54  & 19.54 & 0.680  & 0.807 \\  
		20 & 14.37  & 21.14 & 21.15  & 0.535 & 0.631  & 12.07 & 18.39  & 18.19 & 0.488  & 0.701 \\  
		30 & 13.26  & 19.67 & 19.75  & 0.500 & 0.479  & 11.05 & 17.24  & 16.89 & 0.425  & 0.535 \\  
		40 & 12.41  & 18.20 & 18.40  & 0.375 & 0.364  & 10.36 & 16.10  & 15.65 & 0.377  & 0.396 \\  
		50 & 11.23  & 16.73 & 17.11  & 0.235 & 0.273  &  9.43 & 14.95  & 14.44 & 0.292  & 0.286 \\  
		\rule[-1ex]{0pt}{3.8ex}&& &{\small\textbf{DMSO}} & &&& &{\small\textbf{Water}}&& \\
		10 & 65.47  & 51.65 &   -    & 1.606 & 1.997  & 54.69 & 83.91  & 83.91 & 0.723  & 0.805 \\  
		20 & 57.24  & 47.13 & 47.13  & 1.041 & 1.520  & 53.10 & 80.91  & 80.16 & 0.628  & 0.704 \\  
		30 & 55.59  & 42.61 & 45.86  & 1.072 & 1.323  & 50.86 & 77.91  & 76.57 & 0.517  & 0.602 \\  
		40 & 51.34  & 38.08 & 44.53  & 0.995 & 1.108  & 48.81 & 74.91  & 73.16 & 0.475  & 0.533 \\  
		50 & 49.00  & 33.56 & 43.19  & 0.840 & 0.969  & 47.04 & 71.91  & 69.90 & 0.390  & 0.478 \\
		\hline 
	\end{tabular} 
\end{table*} 

%\begin{figure*}
%	\centering
%	\includegraphics[width=0.8\linewidth]{grafico_tau}
%	\caption{Relaxation time (ps) vs temperature (ºC).	Circles represent experimental values, squares values obtained from MD simulation and crosses those predicted employing the proposed method. \label{fig:graficotau}}
%\end{figure*}

\subsection{Estimation of uncertainties of relaxation times}
\begin{figure}[tbph!]
	\centering
	\includegraphics[width=1.0\linewidth]{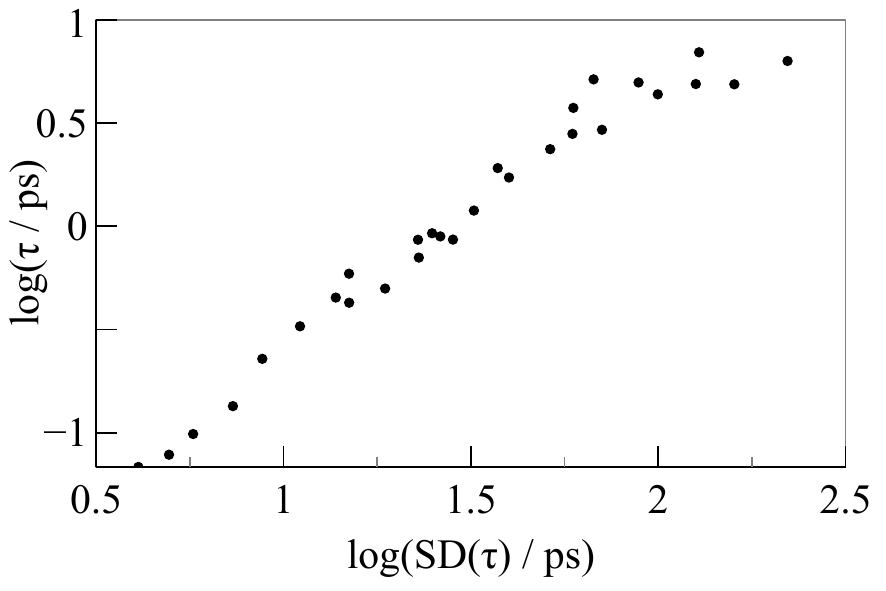}
	\caption[Leyenda corta]{Decimal logarithm of relaxation time / ps vs. decimal logarithm of the estimated standard deviation of relaxation time / ps.}
	\label{fig:errorestau}
\end{figure}

Uncertainties in relaxation times were estimated in the same way that it has be done for permittivities. The computed values can be found in Table \ref{tablatau}, and they are represented in Figure \ref{fig:errorestau}. From them, two conclusions can be drawn. First, the values seems to be in accordance with the statistical noise found for each compound. Second, as stated above, uncertainties tends to increase with relaxation time. Notice that logarithms were used in order to avoid high density of points in some areas of the plot.

\subsection{Temperature dependence}

In order to try our approach, we choose only one experimental value for each compound and pick the one of lowest temperature in each case. By selecting this way we can reach the maximum difference in the temperatures associated to the known value and the one to be predicted, which constitutes a more demanding trial for the proposed method.

Obtaining reasonably converged values for these properties requires very long runs. The total time we used (40 ns) is long enough to reach the corresponding values for chemical species with short relaxation times, like water. The fact that this may not be the case for all the species is easy to circumvent. To this end we model the relaxation times according to Eq. \ref{Modelo activado tau}. That is,

\begin{equation}\label{model tau}
\log \tau(T) = \frac{c_1}{T} + \log T + c_2
\end{equation}

\noindent We used 

\begin{equation}\label{model es}
\varepsilon_s(T) = c_3 + c_4\, T
\end{equation}

\noindent for the relative static dielectric permittivity because it is essentially linear with temperature in the considered range, despite it is well known a more general dependency of the form $a+bT^{-1}$ \cite{hanai1961temperature}. We estimate the $c_k$s parameters of the models from simulation results, and used the expressions above as statistical predictors for the simulation values. 

For relaxation times parameters, we used the ordinary least squares estimator. Instead, weighted least squares was used for permittivities with weight corresponding to the inverse of the variance computed using the proposed method. 

\begin{figure*}[tbph!]
	\centering
	\includegraphics[width=0.8\linewidth]{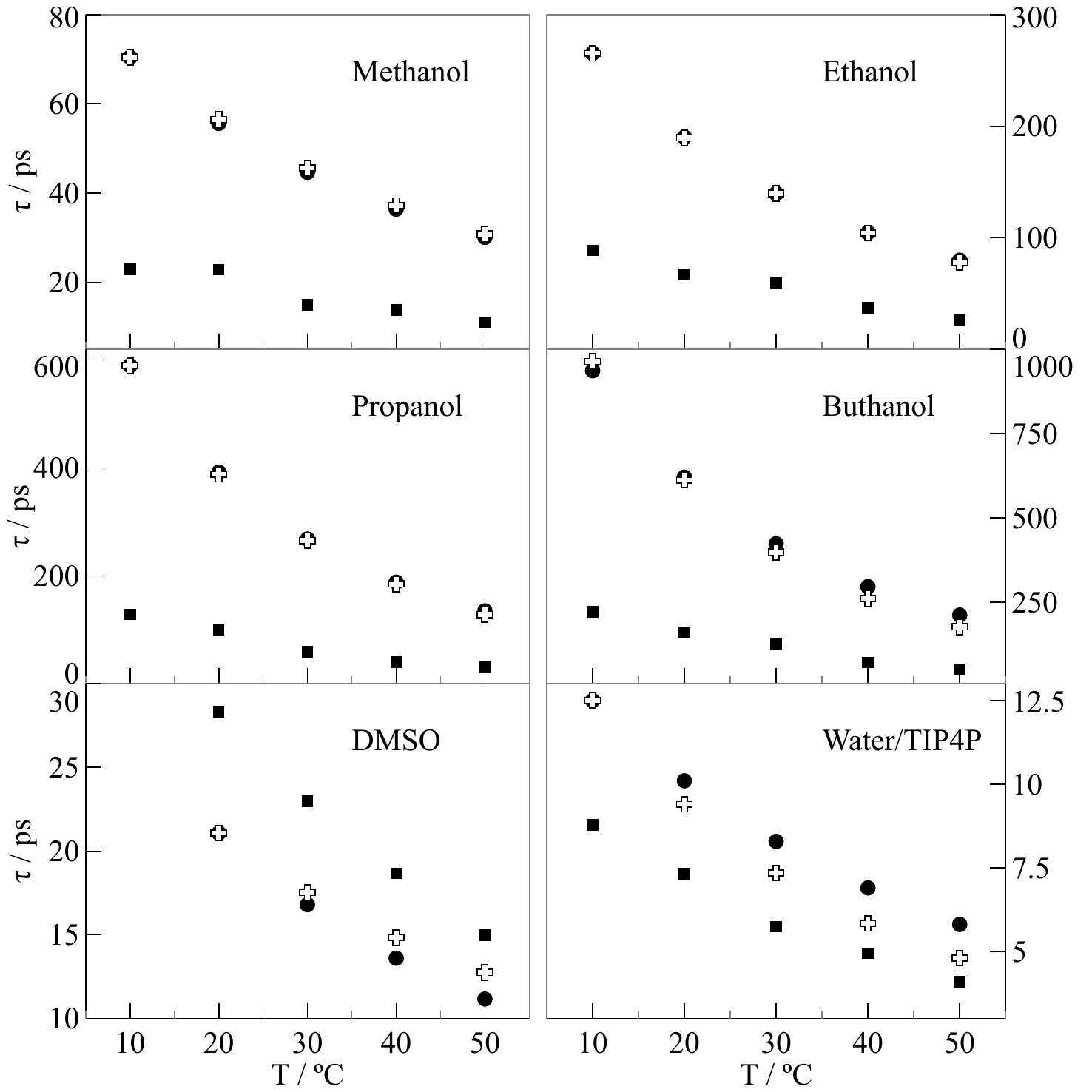}
		\caption{Relaxation time (ps) vs temperature (ºC).	Crosses represent experimental values, squares values obtained from MD simulation and circles those predicted employing the proposed method. \label{fig:graficotau}}
\end{figure*}

\begin{figure*}[tbph!]
	\centering
	\includegraphics[width=0.8\linewidth]{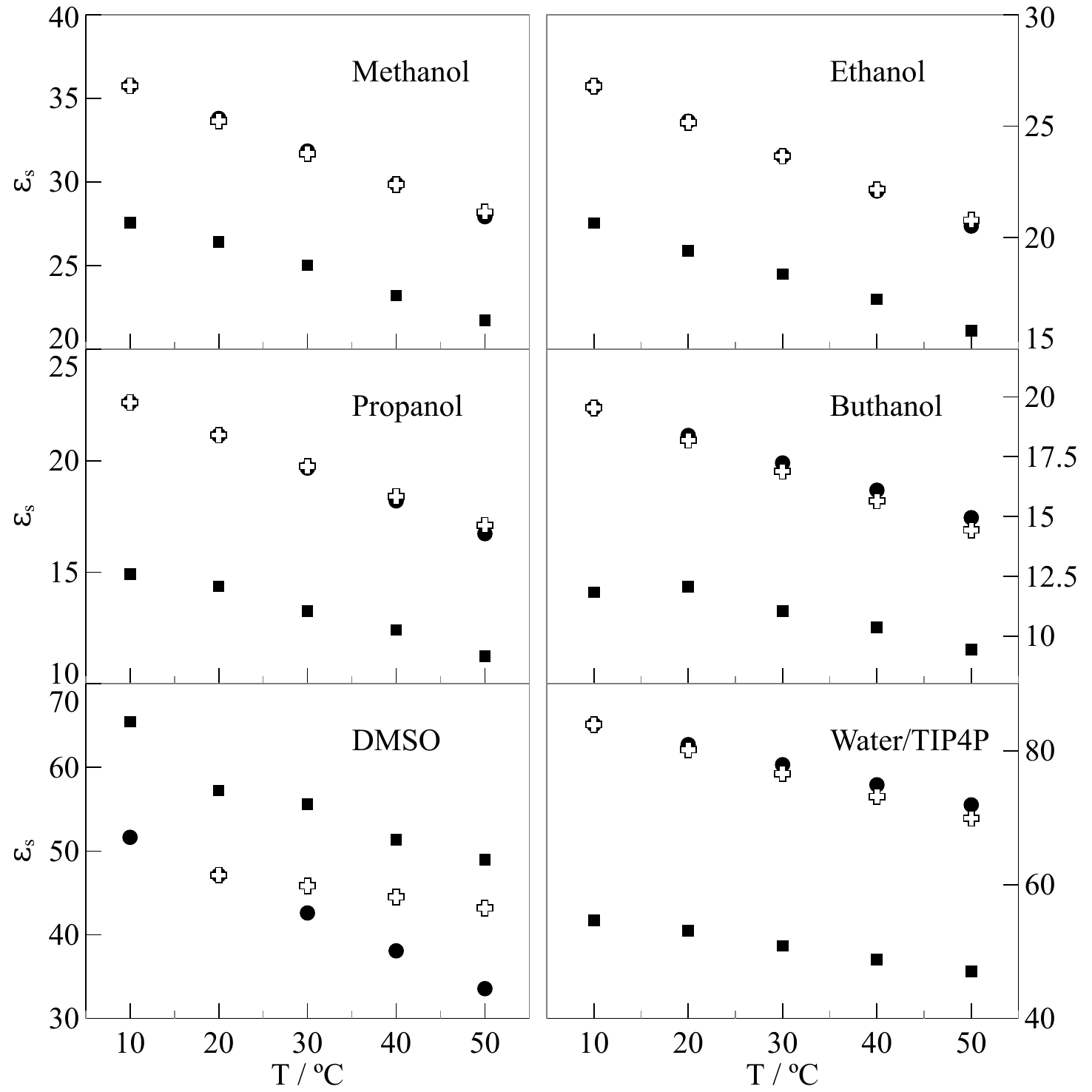}
	\caption[Leyenda corta]{Relative static permittivities vs temperature (ºC).	Crosses represent experimental values, squares values obtained from MD simulation and circles those predicted employing the proposed method.}
	\label{fig:epsilonesestimados}
\end{figure*}

For relaxation times, the experimental, directly computed and final predicted values are compiled in Table \ref{tablatau}.  The predicted results coincide exactly with those from experiments at the lowest temperature because of how we chose the reference values. As expected, the residues tends to increase with temperature. It can be seen by simple inspection of the table \ref{tablatau} that in most cases the proposed method provides  much closer values than the direct calculations. For the convenience of the reader, the results are graphically represented in the figures \ref{fig:graficotau} and \ref{fig:epsilonesestimados}.

\begin{table*}[tbph!]
	\centering\small
	\caption{Magnitudes related to $\tau_D$ (in ps). $^a$ Raw MD results. $^b$ Predicted standard deviation $^c$ Values predicted with the proposed method. $^d$ Experimental values. \label{tablaerroreps}\label{tablatau}}
	\begin{tabular}{ccccccccccccc}
		\hline
		T($^\circ$ C) &Sim.$^a$&SD.$^b$& Pred.$^c$& Exp.$^d$&Sim.$^a$&SD.$^b$& Pred.$^c$& Exp.$^d$&Sim.$^a$&SD.$^b$& Pred.$^c$& Exp.$^d$ \\
		\rule[-1ex]{0pt}{3.8ex}&{\small\textbf{Methanol}} &&& &{\small\textbf{Ethanol}}& & &&{\small\textbf{Propanol}}& & & \\
		10 & 24.92  &0.92& 70.42 & 70.42 & 88.60 &4.97&265.26  & 265.26 & 128.5 &6.95& 589.46 & 589.46\\  
		20 &  22.85 &0.86& 55.60 & 56.44 & 67.21 &5.14& 190.11 & 189.47 & 99.72 &4.35& 392.07 & 388.18\\ 
		30 &  14.96 &0.59& 44.64 & 45.60 & 59.03 &2.80& 139.44 & 139.61 & 59.35 &3.74& 268.19 & 265.26\\ 
		40 &  13.79 &0.45& 36.38 & 37.19 & 37.30 &1.91& 104.43 & 104.02 & 39.97 &1.72& 188.16 & 185.06\\ 
		50 &  11.07 &0.33& 30.06 & 30.78 & 26.21 &0.89& 79.70  &  78.02 & 32.22 &1.19& 135.07 & 128.35\\
		\rule[-1ex]{0pt}{3.8ex}&{\small\textbf{Buthanol}} &&& &{\small\textbf{DMSO}}& & &&{\small\textbf{Water}}& & & \\		
		10 & 221.33 &6.31& 936.21 & 936.21 &  -   &- &   -    &  -     & 8.78 &0.23& 12.50 & 12.50 \\
		20 & 159.58 &4.87& 620.49 & 612.13 & 28.33&0.86& 21.08 & 21.08 & 7.33 &0.13& 10.10 & 9.40  \\
		30 & 125.97 &4.88& 423.02 & 397.89 & 22.97&0.70& 16.80 & 17.53 & 5.74 &0.10& 8.29  & 7.35  \\
		40 &  70.77 &2.93& 295.86 & 260.91 & 18.67&0.50& 13.60 & 14.83 & 4.95 &0.08& 6.90  & 5.84  \\
		50 &  51.48	&2.36& 211.76 & 176.84 & 14.97&0.43& 11.16 & 12.75 & 4.10 &0.07& 5.81  & 4.80  \\
		\hline 
	\end{tabular} 
\end{table*} 

In the case of water, the improvement of our approach over direct calculation decreases with temperature until the last point in which the latter turns to be the most accurate. This behavior is due to direct calculation turns to be very accurate for this system in such conditions. It is worth noting that if we had chosen the highest temperature value for reference, our approach would outperform the direct calculation in all cases. Furthermore, the relative improvement would substantially increase. The most likely scenario is one in between as the measurement are commonly carried out at standard temperature. 

Results for relative static dielectric permittivity can be found in Table \ref{tablaerroreps} and Figure \ref{fig:epsilonesestimados} provides a graphical representation of them. Our estimation greatly improves upon MD results except for DMSO. As by definition our method is exact at the reference temperature, faithful comparisons should be made far from it. At 50$^\circ$C, the absolute value of the quotient between relative errors of our method and raw MD results are 0.04, 0.05, 0.06, 1.66 and 0.08, for methanol, ethanol, propanol, buthanol, DMSO and water, respectively. 

While in most cases our method decrease relative error in a factor between 12 and 25, it increased the error for DMSO. It is due to MD using OPLS provides less accurate temperature relative variation for this compound.

We underline that modeling the simulated results as we did mitigates the issues related to lack of convergence of single calculated values.

\section{Summary}
The two main issues addressed in this work can be summarized as follows. %are summarized bellow.
 
Firstly, relative static dielectric permittivity calculated through MD is a random variable. We proposed a method for predicting its uncertainty taking into account that the total dipole moment of the system is an auto-correlated variable. This method has a smoother dependency with temperature that a simpler numerical approach,  which suggests that the former is more reliable.

Secondly, for most substances no experimental data at different temperatures is available for relaxation times nor static dielectric constants. Owing to this absence, in this work, a method for predicting these magnitudes was proposed. The latter requires the usage of one known experimental value, and at least two MD simulation. It is based in the general idea that it is easier predict temperature dependencies, than absolute values depending upon many more variables.

The procedure consists in predict the value of this magnitude by means of MD at the temperature of interest and the one at which the measurement was performed, compute their ratio and multiply it with the measured value. In the case of relaxation times, we derived this relationship from theoretical considerations.

In neither case the directly computed values through MD were used. Instead, we modeled them according to equations \ref{model tau} and \ref{model es}, and estimated the models parameters. That way we avoided convergence issues due to limited simulation time.  

In most cases, a huge improvement upon raw MD results was found. In few cases MD raw results were a little more accurate.

\section{Acknowledgements}	
Support of this work by Consejo Nacional de Investigaciones Cient\'{\i}ficas y
T\'{e}cnicas of Argentina and ANPCyT (Argentina) through PICT-2016-2303 is greatly appreciated. R.M.I. and C.M.C. are members of CONICET.

\bibliographystyle{unsrt}
\bibliography{refs}

\end{document}